\documentclass[fleqn,twoside]{article}
\usepackage{espcrc2}

\usepackage{graphicx}
\usepackage{epsfig}
\usepackage[figuresright]{rotating}

\newcommand{\AmS}{{\protect\the\textfont2
  A\kern-.1667em\lower.5ex\hbox{M}\kern-.125emS}}
\newcommand{\drhoe}{D^0\to \rho^-e^+\nu_e}
\newcommand{\dkste}{D^0\to K^{*-}e^+\nu_e}
\newcommand{\dke}{D^0\to K^-e^+\nu_e}
\newcommand{\dpie}{D^0\to \pi^-e^+\nu_e}

\newcommand{\dkpp}{D^0\to K^-\pi^+\pi^0}

\newcommand{\dkpppc}{D^0\to K^-\pi^+\pi^+\pi^-}

\newcommand{\dpkpp}{D^+\to K^-\pi^+\pi^+}

\newcommand{\dpksp}{D^+\to K_S\pi^+}

\newcommand{\vcs}{V_{cs}}
\newcommand{\vcd}{V_{cd}}
\newcommand{\vub}{V_{ub}}

\newcommand{\vtd}{V_{td}}

\newcommand{\vts}{V_{ts}}
\newcommand{\vcb}{V_{cb}}
\newcommand{\vtb}{V_{tb}}

\title{An Experimenter's View of Lattice QCD}

\author{I. Shipsey
\address{Department of Physics, Purdue University,
        West Lafayette, IN 47907, U.S.A.}
        }

\begin{document}

\begin{abstract}
Lattice QCD has the potential this decade to maximize the
sensitivity of the entire flavor physics program to new physics
and pave the way for understanding physics beyond the Standard
Model at the LHC in the coming decade. However, the challenge for
the Lattice is to demonstrate reliability at the level of a few
per cent given a past history of 10-20\% errors. The CLEO-c
program at the Cornell Electron Storage Ring is providing the data
that will make the demonstration possible. \vspace{1pc}
\vskip -1.5em
\end{abstract}

\maketitle

\section{INTRODUCTION}

Lattice QCD (LQCD) is  the only complete definition of
perturbative and non-perturbative QCD, but is also a technique
with a history of results that deviate from experiment by 10-20\%.
This is beginning to change. Recent advances in LQCD culminated in
the precision calculations of nine, previously measured, diverse
quantities~\cite{Davies}, that agree with experiment within a
few per cent. This could not have come at a better time, as the
era of experimental precision quark flavor physics we are now in,
depends crucially on the precise calculation of non-perturbative
quantities in the beauty sector. How will the community know if
the lattice calculations of these quantities are correct? Charm at
threshold can provide the data necessary to test the calculations,
and an experiment operating there, CLEO-c,  has just begun.

%\section{THE BIG QUESTIONS IN FLAVOR PHYSICS}
\subsection{Big Questions in Flavor Physics}

The big questions in quark flavor physics are: (1) ``What is the
dynamics of flavor?'' The gauge forces of the standard model (SM)
do not distinguish between fermions in different generations. The
electron, muon and tau all have the same electric charge, quarks
of different generations have the same color charge. Why
generations? Why three? (2) ``What is the origin of
baryogenesis?'' Sakharov gave three criteria, one is
$CP$-violation~\cite{Sakharov}. There are only three known
examples of $CP$-violation: the Universe, and the beauty and kaon
sectors. However, SM $CP$-violation is too small, by many orders
of magnitude, to give rise to the baryon asymmetry of the
Universe. Additional sources of $CP$-violation are needed. (3)
``What is the connection between flavor physics and electroweak
symmetry breaking?''
Extensions of the SM, for example
supersymmetry, contain flavor and $CP$-violating couplings that
should show up at some level in flavor physics but precision
measurements and precision theory are required to detect the new
physics.

%\section{FLAVOR PHYSICS TODAY}
\subsection{Flavor Physics Today}

This is the decade of precision flavor physics. In the
``$\sin 2 \beta$ era'', the goal is to over-constrain the CKM
matrix with a range of measurements in the quark flavor changing
sector of the SM at the per cent level. If inconsistencies are
found between, for example, measurements of the sides and angles
of the $B_d$ unitarity triangle, it will be evidence for new
physics. Many experiments will contribute including BaBar and
Belle, CDF, D0, and BTeV at Fermilab, ATLAS, CMS, and LHC-b at the
LHC, CLEO-c, and experiments studying rare kaon decays.

\begin{figure*}[btp]
%\centerline{
\includegraphics[width=0.47\textwidth]{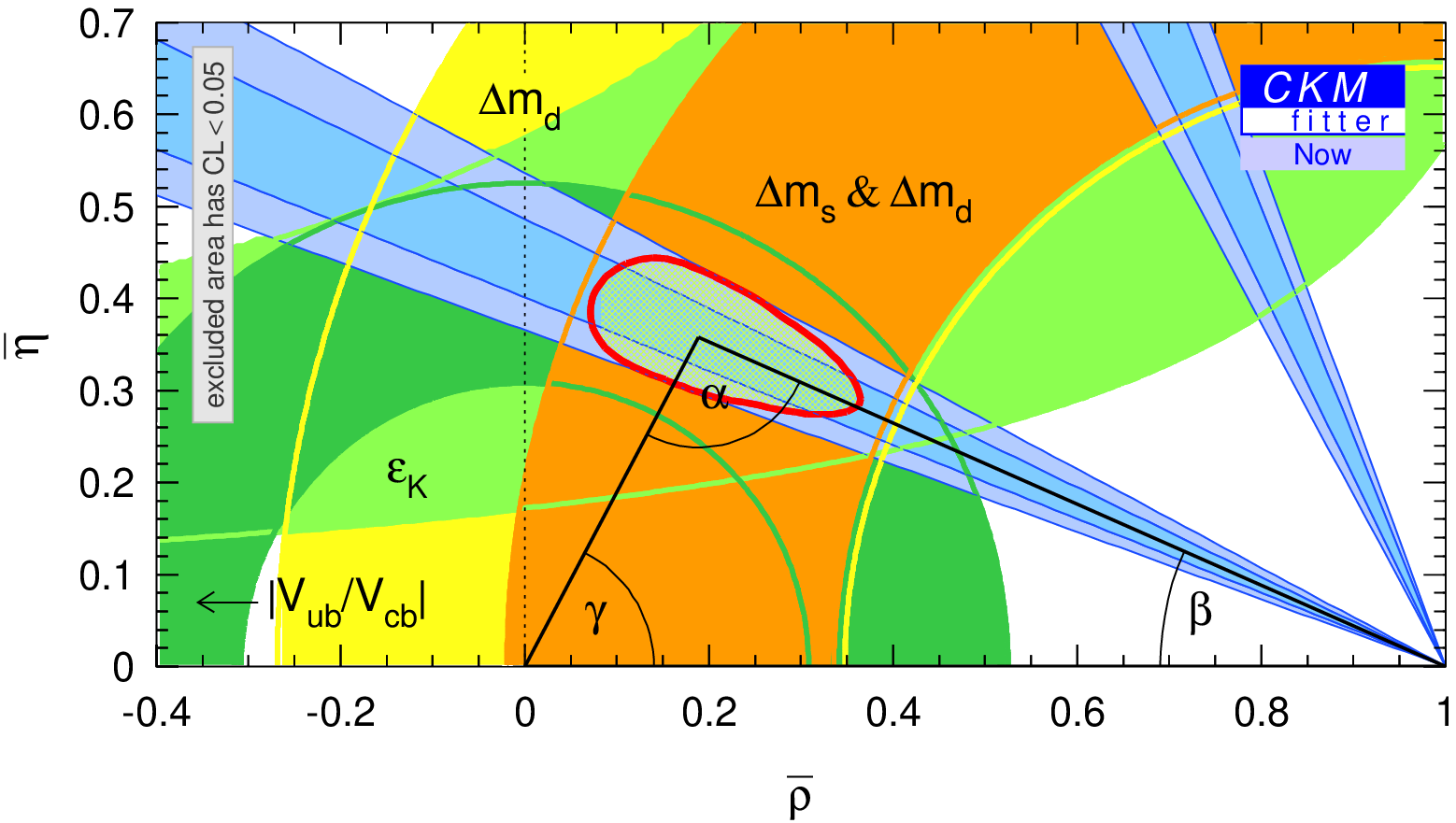}
\hfill
\includegraphics[width=0.47\textwidth]{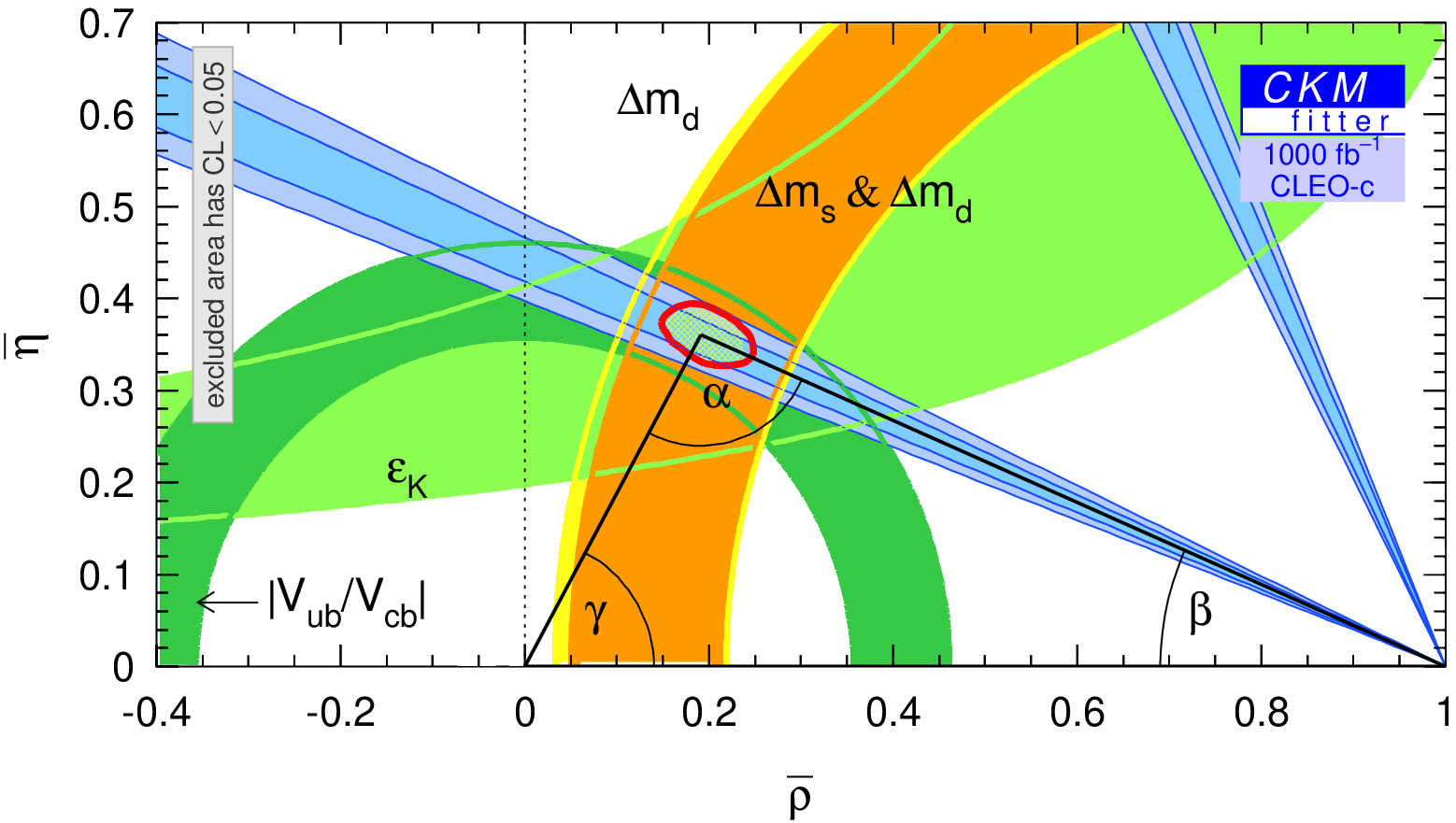}
%\epsfig{figure=CKM_fitter_2004_theory2percentaug12.eps,height=0.27\textwidth}
%}
\vskip -2em
\caption{Lattice impact on the $B_d$ unitarity
triangle from $B_d$ and $B_s$ mixing, $|\vub| / |\vcb|$,
$\epsilon_K$, and $\sin 2 \beta$.
(a) Summer 2004 status of the constraints.
(b) Prospects under the assumption that LQCD
calculations of $B$ system decay constants and semileptonic form
factors achieve the projections in Table~\ref{table:combine}.}
\label{CKM}
\vskip -0.5em
\end{figure*}

However, the study of weak interaction phenomena, and the
extraction of quark mixing matrix parameters remain limited by our
capacity to deal with non-perturbative strong interaction
dynamics.
Current constrains on the CKM matrix are shown in Fig.~\ref{CKM}(a).
The widths of the constraints, except that of $\sin 2 \beta$,
are dominated by the error bars on the calculation of
hadronic matrix elements. Techniques such as lattice QCD directly
address strongly coupled theories and have the potential to
eventually determine our progress in many areas of particle
physics. Recent advances in LQCD have produced calculations of
non-perturbative quantities such as $f_\pi$, $f_K$, and heavy
quarkonia mass splittings that agree with
experiment~\cite{Davies}.
%with
%accuracies in the 10-20\% level for systems involving one or two
%heavy quark such as $B$ and $D$ mesons, and $\psi$ and $\Upsilon$
%quarkonia. Several per cent precision has been reached in certain
%golden quantities~\cite{Davies}
Several per cent precision in charm and beauty decay constants and
form factors is hoped for, but the path to higher precision is
hampered by the absence of accurate charm data against which to
test lattice techniques.

%\section{CLEO-c AND THE LATTICE}
\subsection{CLEO-c and the Lattice}

To meet this challenge the CLEO collaboration has converted CLEO
and CESR into a charm and QCD factory operating at charm threshold
where the experimental conditions are
optimal~\cite{CLEO-cyellowbook}. In a pilot run in 2003 CLEO-c
recorded a data sample about one fiftieth of design
%. The data has
%exceptionally low background and
that has already allowed the most precise measurements of several
quantities that are important tests of LQCD including $f_{D^+}$
and ${\cal B}(D^0 \rightarrow \pi^- e^+ \nu_e)$ or are important
to set the scale for heavy quark physics including ${\cal B}(D^+
\rightarrow K^- \pi^+ \pi^+)$. Beginning September 2004 CLEO-c
will obtain charm data samples one to two orders of magnitude
larger than any previous experiment.
%operating in this energy
%range, and with a detector that is significantly more powerful
%than any previous detector to operate at charm threshold.
This data has the potential to provide unique and crucial tests of
LQCD with accuracies of 1-2\%.

If LQCD passes the CLEO-c test,
%CLEO-c data combined with LQCD
%predictions will lead to a significant improvement in our
%knowledge of the quark couplings in the charm sector. Moreover,
the community will have much greater confidence in lattice
calculations of decay constants and semileptonic form factors in
beauty physics. When these calculations are combined with
500~fb$^{-1}$ of $B$ factory data, and improvement in the direct
measurement of $|\vtb|$ expected from the Tevatron
experiments~\cite{Swain}, they will allow a significant reduction
in the size of the errors on the quark couplings $|\vub|, |\vcb|,
|\vtd| {\rm ~and~} |\vts|$, quantitatively  and qualitatively
transforming knowledge of the $B_d$ unitarity triangle, see
Fig.~\ref{CKM}(b), and thereby maximizing the sensitivity of
heavy quark physics to new physics.
%The combination of LQCD and CLEO-c is the
%missing piece in the quest to understand the origin of CP
%violation and quark mixing. LQCD and CLEO-c together will enable
%the particle physics community to draw back the curtain of
%hadronic uncertainty that has blocked the view for 40 years,  and
%see clearly through the heavy quark window to the new physics that
%lies beyond the SM.

Of equal importance, LQCD combined with CLEO-c allows a
significant advance in understanding and control over
strongly-coupled, non-perturbative quantum field theories in
general. Field theory is generic, but weak coupling is not. Two of
the three known interactions are strongly coupled: QCD and gravity
(string theory). An understanding of strongly coupled theories may
well be a crucial element in helping to interpret new phenomena at
the
%LHC and ILC
high energy frontier.

%\section{KEY CHARM FLAVOR PHYSICS TESTS OF LQCD}
\section{TESTS OF LQCD WITH CHARM}

\subsection{Decay Constants}

The $B_d$ $(B_s)$ meson mixing probability can be used to
determine $ |V_{td}|$ $(|V_{ts}|)$.
\begin{equation}
\Delta m_d \propto |V_{tb}V_{td}|^2 f_{B_d}^2 B_{B_d}
\end{equation}
The $B_d$ mixing rate is measured with exquisite precision
(1.4\%)~\cite{PDG2004} but the decay constant is calculated with a
precision of about 15\%. If theoretical precision could be
improved to 3\%, $|V_{td}|$ would be known to about 5\% without
any need for improvement in the experimental measurement.

Since LQCD hopes to predict $f_B/f_{D^+}$ with a small error,
measuring $f_{D^+}$ would allow a precision prediction for $f_B$.
Hence a precision extraction of $|V_{td}|$ from the $B_d$ mixing
rate becomes possible. Similar considerations apply to $B_s$
mixing  once it is measured i.e. a precise determination of
$f_{D_s^+}$ would allow a precision prediction for $f_{B_s}$ and
consequently  a precision measurement of $|\vts|$. Finally the
ratio of the two neutral $B$ meson mixing rates determines $|\vtd| /
|\vts|$, but $|\vts| = |\vcb|$ by unitarity and $|\vcb|$ is known
to a few per cent, and so the ratio determines $\vtd$. Which
method of determining $|\vtd|$ will have the greater utility
depends on which combination of hadronic matrix elements have the
smallest error.

Charm leptonic decays can be used to measure the charm decay
constants $f_{D_s^+}$ and  $f_{D^+}$ because $|V_{cs}|$ and
$|V_{cd}|$ are known from unitarity to 0.1\% and 1\% respectively.
\begin{equation}
{ {{\cal B}(D^+ \rightarrow \mu \nu_\mu) }\over {\tau_{D^+} } }=
{\rm (const.)} f_{D^+}^2 |V_{cd}|^2
\end{equation}
(Charge conjugation is implied throughout this paper.) The
measurements also provide a precision test of the lattice
calculations of $f_{D_s^+}$ and  $f_{D^+}$. At the start of 2004
$f_{D^+}$ was experimentally undetermined and $f_{D_s^+}$ was
known to 33\%.

\subsection{Semileptonic form factors}

$V_{ub}$ measures the length of the side opposite the angle
$\beta$ in the $B_d$ unitarity triangle and consequently it is a
powerful check of the consistency of the CKM matrix paradigm of
$CP$-violation. $|V_{ub}|$ is determined from beauty semileptonic
decay
\begin{equation}
{{d\Gamma(B \rightarrow \pi e^- \bar\nu_e)} \over {dq^2}} = { \rm
(const.)} |V_{ub}|^2f_+(q^2)^2
\label{eq:Bsemi}
\end{equation}
The differential rate depends on a form factor, $f_+(q^2)$ that
parameterizes the strong interaction non-perturbative effects. A
recent representative value of $|\vub|$ determined from $B
\rightarrow  \pi \ell^- \bar{\nu_e}$ is~\cite{Ali}:
\begin{equation}
|V_{ub}| = (3.27 \pm 0.70 \pm 0.22^{+0.85}_{-0.51}) \times 10^{-3}
\end{equation}
where the uncertainties are experimental statistical and
systematic, and from the LQCD calculation of the form factor,
respectively.
%The experimental errors, now large, are expected to
%be reduced to 5\% with a $B$ factory data sample of $500 {\rm fb
%^{-1}}$ per experiment, and the theory error will dominate.
The large experimental errors are expected to
be reduced to 5\% with $B$ factory data samples of
$500 {\rm fb}^{-1}$ each, and the theory error will dominate.

%The differential charm semileptonic rate
%\begin{equation}
%{{d\Gamma(D \rightarrow \pi e^+ \nu_e)} \over {dq^2}} ={\rm
%(const.)} |V_{cd}|^2f_+(q^2)^2
%\end{equation}
%tests calculations of charm semileptonic form factors because the
%charm CKM matrix elements are known from unitarity.  A  precision
%measurement of the $D \rightarrow \pi$ form factor tests the LQCD
%calculation of the $D \rightarrow \pi$ form factor.
Again, because the charm CKM matrix elements are know from unitarity,
the differential charm semileptonic rate
\begin{equation}
{{d\Gamma(D \rightarrow \pi e^+ \nu_e)} \over {dq^2}} ={\rm
(const.)} |V_{cd}|^2f_+(q^2)^2
\end{equation}
tests calculations of charm semileptonic form factors.
Thus, a precision measurement tests the LQCD
calculation of the $D \rightarrow \pi$ form factor.
As the form
factors governing $B \rightarrow  \pi e^- \bar{\nu_e}$ and $D
\rightarrow \pi e^+ \nu_e$ are related by heavy quark symmetry,
the charm test gives confidence in the accuracy of  the $B
\rightarrow \pi$ calculation. The $B$ factories can then use a
tested LQCD prediction of the $B \rightarrow \pi$ form factor to
extract a precise value of $|V_{ub}|$ from Eq.~(\ref{eq:Bsemi}).
%a measurement of $d\Gamma(B \rightarrow \pi e^- \bar\nu_e) / dp_{\pi}$.
At the start of 2004,
${\cal B} (D \rightarrow \pi e^+ \nu_e)$ had been determined to
45\%~\cite{PDG2004,PDG2004_C}, but the absolute value of the $D
\rightarrow \pi$ form factor had not been measured.

\section{FIRST RESULTS FROM CESR-c AND CLEO-c}

The Cornell Electron Storage Ring (CESR) has been upgraded to
CESR-c with the installation of 12 wiggler magnets to increase
damping at low energies. Six wigglers were installed in the summer
of 2003 and the remainder this summer. Between September 2003 and
March 2004 a CLEO-c pilot run accumulated $57.1~{\rm pb}^{-1}$ at the
$\psi(3770)$, about three times larger than any previous sample
collected at this energy. The accelerator achieved a luminosity of
$L = 4.6\times 10^{31}~{\rm cm}^{-2}{\rm s}^{-1}$, as anticipated.
Starting in September 2004 CLEO-c will take data at
$\sqrt{s} \sim 3770$~MeV,
$\sqrt{s} \sim 4140$~MeV, and
$\sqrt{s} \sim 3100$~MeV ($J/\psi$).
The design luminosity at these energies
ranges from $5 \times 10^{32}~{\rm cm^{-2} s^{-1}}$ down to about
$1 \times 10^{32}~{\rm cm^{-2} s^{-1}}$ yielding 3~fb$^{-1}$ each
at the $\psi^{\prime\prime}$ and at $\sqrt{s} \sim 4140$~MeV above
$D_s \bar{D_s}$ threshold, and 1~fb$^{-1}$ at the $J/\psi$ in a
Snowmass year of $10^7~{\rm s}$. These integrated luminosities
correspond to samples of 20 million $D \bar{D}$ pairs, 1.5 million
$D_s \bar{D_s}$ pairs, and one billion $J/\psi$
decays~\cite{CLEO-cyellowbook}. These datasets will exceed those
of the BESII (Mark III) experiment by factors of 130 (480), 110
(310) and 20 (170), respectively.

The CLEO-c detector is a minimal modification of the well
understood CLEO III detector. A silicon vertex detector was
replaced with a small-radius low-mass drift chamber, and the
magnetic field was lowered to 1.0~T from $1.5~{\rm T}$.
CLEO-c is the first
modern detector to operate at charm threshold.

\subsection{Analysis Technique }

\begin{figure}[btp]
\centerline{\epsfig{figure=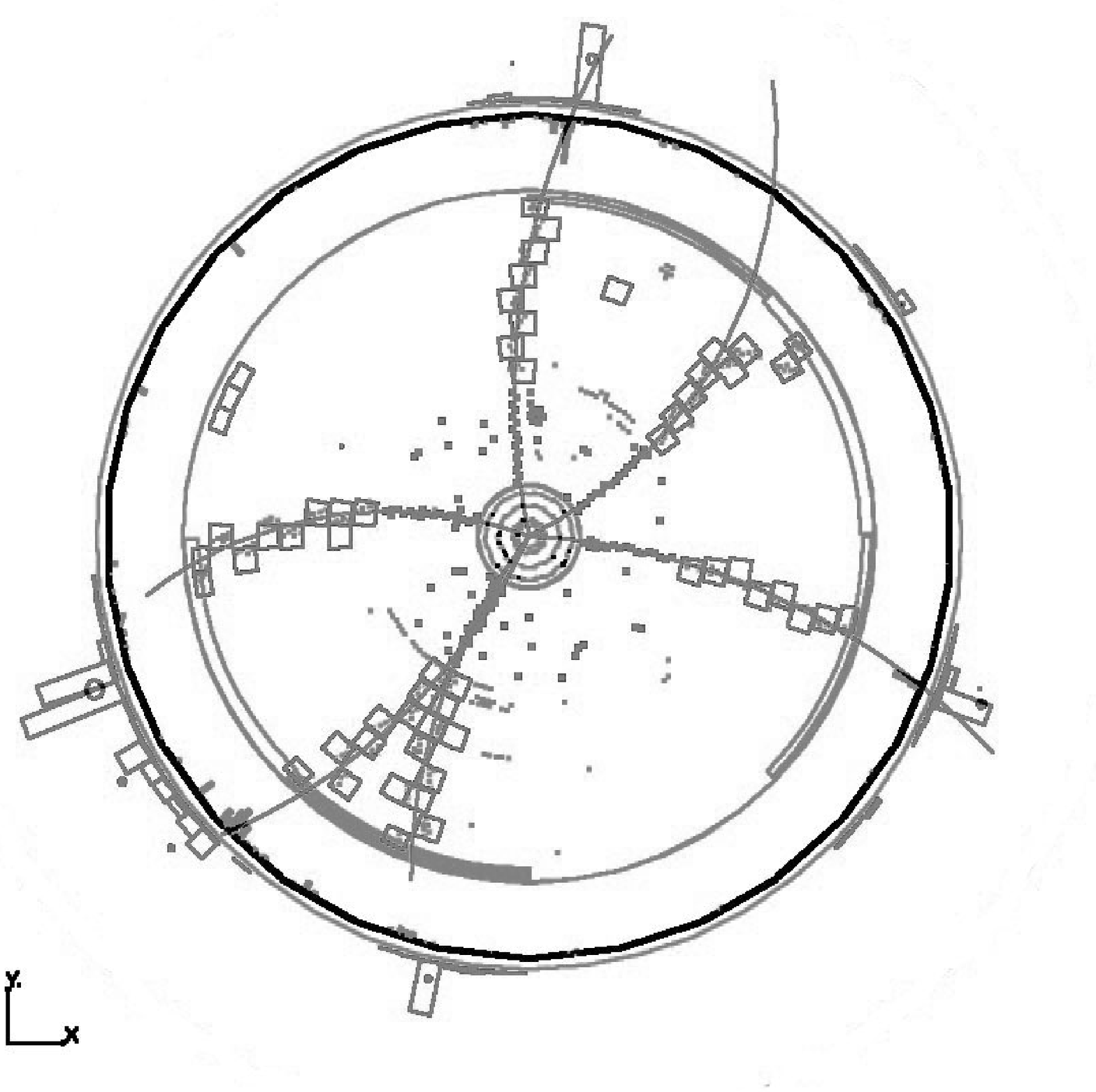,width=0.45\textwidth}}
%\centerline{\epsfig{figure=kpipi_kpipi.eps,width=0.45\textwidth}}
\vskip -1.0em \caption{ A CLEO-c event where $D^+ \rightarrow K^-
\pi^+ \pi^+, D^- \rightarrow K^+ \pi^- \pi^-$.} \vskip -1.5em
\label{event_hadronic}
\end{figure}

There are significant advantages to running at charm threshold. As
$\psi \rightarrow D \bar D$, the strategy is to fully
reconstruct one $D$ meson in a hadronic final state, which is
referred to as the tag, and then to analyze the decay of the
second $D$ meson in the event to extract inclusive or exclusive
properties. A typical event, in which both $D$ mesons have been
reconstructed, is shown in Fig.~\ref{event_hadronic}.

\begin{figure}[btp]
\centerline{\epsfig{figure=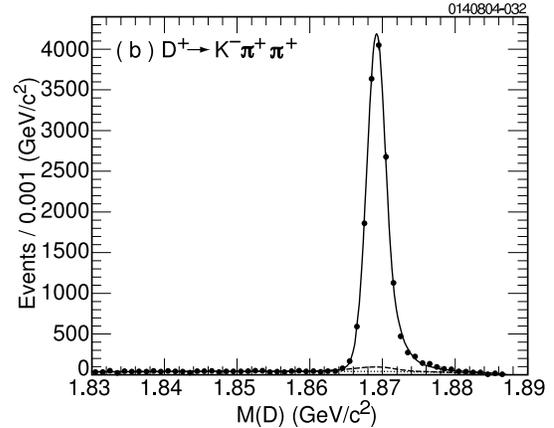,width=0.45\textwidth}}
\vskip -2.0em
\caption{Distribution of calculated $M(D)$ values for single tag
$D$ candidates in the mode $D^+ \rightarrow K^- \pi^+ \pi^+$.
Preliminary. } \label{single}
\vskip -1.5em
\end{figure}

As $E_{\rm beam}=E_D$, a requirement that the candidate have energy
close to the beam energy is made, and the beam-constrained
candidate mass,
$M(D)  = \sqrt{E_{{\rm beam}}^2 - p_{{\rm cand}}^2}$,
is computed. The $M(D)$ distribution for the mode $D^+
\rightarrow K^- \pi^+ \pi^+$ is shown in Fig.~\ref{single}.
%Charm events
%produced at threshold are extremely clean.
The signal to noise,
which is optimal at threshold, is about 50:1.

Charm mesons have many large branching ratios to low multiplicity
final states. In consequence the tagging efficiency is very high,
about 25\%, this should be compared to less than 1\% for $B$
tagging at a $B$ factory.

\begin{figure}[btp]
\centerline{\epsfig{figure=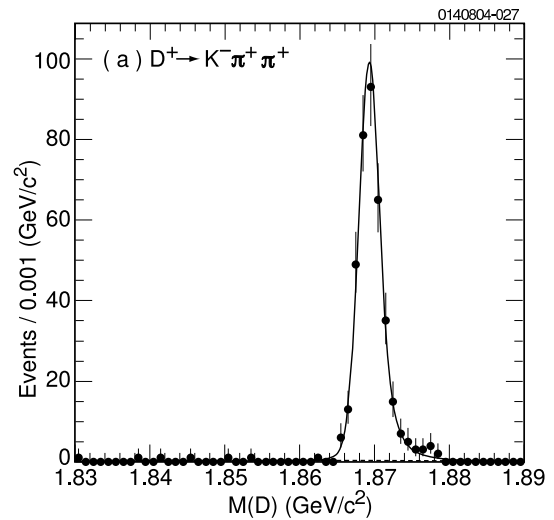,width=0.45\textwidth}}
\vskip -2.0em
\caption{Projection of the double tag $D^+D^-$ candidate masses
onto the $M(D)$ axis for $D^+ \rightarrow K^- \pi^+ \pi^+, D^-
\rightarrow K^+ \pi^- \pi^-$. Preliminary.} \label{double}
\vskip -1.5em
\end{figure}

Tagging creates a single $D$ meson beam of known momentum. This is
a particularly favorable experimental situation.
Figure~\ref{double} shows $M(D)$ of the second $D$ meson in events
where both $D$ mesons have decayed into the $K^- \pi^+ \pi^+$
final state.
%These double tag events, which are key to making
%absolute branching fraction measurements, are pristine.
%The absolute branching ratio is computed as
These double tag events are pristine.
They are key to making absolute branching fraction measurements:
\begin{equation}
{\cal B} (D^+ \rightarrow K^- \pi^+ \pi^+) = { {N(K^- \pi^+
\pi^+)}\over { \epsilon (K^- \pi^+ \pi^+) \times N(D^-)}}
\end{equation}
where $N(K^- \pi^+ \pi^+)$ is the number of $D^+ \rightarrow K^-
\pi^+ \pi^+$ observed in tagged events, $\epsilon (K^- \pi^+
\pi^+)$ is the reconstruction efficiency and $N(D^-)$ is the
number of tagged events. In a method similar to that pioneered by
Mark III~\cite{Balt,Adler}, CLEO fits to the observed single
tag and double tag yields for five $D^+$ and $D^0$ modes, and
finds the preliminary branching ratios listed in
Table~\ref{table:br}. The statistical errors are comparable to
previous measurements, while the preliminary systematic errors are
likely to be reduced in the near future. This is the most precise
measurement of ${ \cal B} (\dpkpp) $.
\begin{table}[tbp]
\caption{Preliminary CLEO-c absolute charm branching ratios.
Further detail in Ref.~\cite{CLEO_had}.}
 \label{table:br}
\begin{tabular}{@{}ll}
\hline
Mode                      & ${\cal B}$ (\%)  \\
\hline
$D^0 \rightarrow K^- \pi^+$    &  $3.92 \pm 0.08 \pm 0.23$  \\
$\dkpp $                     &  $14.3 \pm 0.3 \pm 1.0 $ \\
$\dkpppc $                    & $8.1 \pm 0.2 \pm 0.9$    \\
$\dpkpp $                    &  $9.8 \pm 0.4 \pm 0.8$  \\
$ \dpksp $                    & $1.61 \pm 0.08 \pm 0.15$   \\
 \hline
\end{tabular}
\vskip -1.5em
\end{table}

The fit also returns the number of $D$ meson pairs, from which the
cross section is obtained:
\begin{equation}
\sigma(e^+ e^- \rightarrow D \bar D) =
(6.06 \pm 0.13 \pm 0.32)~{\rm nb}
\end{equation}
where the uncertainties are statistical and systematic,
respectively. The cross section is independent of charm branching
ratios.

The CLEO-c $\psi(3770)$ integrated luminosity goal of $3~{\rm
fb^{-1}}$ may sound small compared to the $500~{\rm fb^{-1}}$
expected at each of the $B$ factories.
The ability to
perform a tagged analysis is comparable at the two facilities, however,
because the tagging efficiency is about 25 times larger at a charm
factory than at a $B$ factory, and the cross section is about six
times larger. Hence,
%\begin{eqnarray}
%{ N(B~{\rm tags~at~a~}B~{\rm factory}) \over
  %N(D~{\rm tags~at~a~charm~factory})   }
 %= \non\\
%{ \sigma(B\bar B) \times \epsilon({\rm tag})
    %(\int L dt = 500~{\rm fb}^{-1} ) \over
  %\sigma(D\bar D) \times \epsilon({\rm tag})
    %(\int L dt =   3~{\rm fb}^{-1} ) } \sim 1.
%\end{eqnarray}
\begin{equation}
{ N(B~{\rm tags~at~a~}B~{\rm factory}) \over
  N(D~{\rm tags~at~a~charm~factory})   } \sim 1.
\end{equation}

The absolute branching ratios ${\cal B}(\dpkpp)$, ${\cal B}(D^0
\rightarrow K^-\pi^+)$, and ${\cal B} (D_s^+ \rightarrow \phi
\pi^+ )$ are important as, currently, all other $D^+$, $D^0$ and
$D_s^+$ branching ratios are determined from ratios to one or the
other of these branching fractions~\cite{PDG2004}. In consequence,
nearly all branching fractions in the $B$ and $D$ sectors depend
on these reference modes.  Projections for the expected precision
with which the reference branching ratios will be measured with
the full CLEO-c data set are given in Table~\ref{table:brproj}.
CLEO-c will set the scale for all heavy quark measurements.
\begin{table}[tbp]
\caption{CLEO-c hadronic branching ratio projections.
Further detail in Ref.~\cite{CLEO-cyellowbook}.}
 \label{table:brproj}
\begin{tabular}{@{}lll}
\hline
Mode                      & \multicolumn{2} {c} { $ \delta {\cal B} / {\cal B} $ (\%) } \\
& PDG 2004 & CLEO-c  \\ \hline
$D^0 \rightarrow K^- \pi^+$    & 2.4\%  & 0.6\%   \\
$\dpkpp $                        &  6.1\%  &  0.7\% \\
$D_s^+ \rightarrow \phi \pi $                    & 12.5\%~\cite{BaBar}  & 1.9\%   \\
 \hline
\end{tabular}
\vskip -1.5em
\end{table}

\subsection{Measurement of the Charm Decay Constant }

\begin{figure}[btp]
\centerline{\epsfig{figure=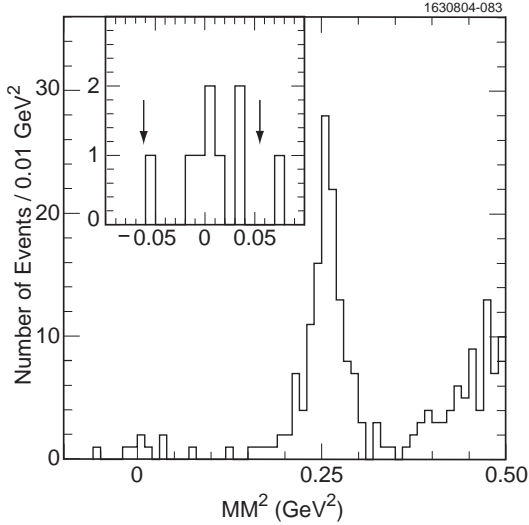,width=0.45\textwidth}}
\vskip -1.0em
\caption{ The $MM^2$ distribution in events with $D^-$ tag, a
single charged track of the correct sign, and no additional
(energetic) showers. The insert shows the signal region for $D^+
\rightarrow \mu \nu_\mu$.
A $\pm 2 \sigma$ range is
indicated by the arrows.
Preliminary. } \label{missmass}
\vskip -1.5em
\end{figure}

The measurement of the leptonic decay $D^+ \rightarrow \mu^+
\nu_\mu$ benefits from the fully tagged $D^-$ at the $\psi(3770)$.
One observes a single charged track recoiling against the tag that
is consistent with a muon of the correct sign. Energetic
electromagnetic showers un-associated with the tag are not
allowed. The missing mass $MM^2 = m_{\nu}^2$ is computed; it %, and
peaks at zero for a decay where only a neutrino is unobserved. The
clear definition of the initial state, the cleanliness of the tag
reconstruction, and the absence of additional fragmentation tracks
make this measurement straightforward and nearly background-free.
The $MM^2$ distribution is shown in Fig.~\ref{missmass}.
There are 8 candidate signal events, and $1.07 \pm 1.07$ background
events. After correcting for efficiency, CLEO-c finds
\begin{equation}
{\cal B} (D^+ \rightarrow \mu^+ \nu_\mu) = (3.5 \pm 1.4 \pm 0.6)
\times 10^{-4},
\end{equation}
where the uncertainties are statistical and systematic,
respectively. Under the assumption of three generation unitarity,
and using the precisely known $D^+$ lifetime, CLEO-c obtains
\begin{equation}
f_{D^+} = (201 \pm 41 \pm 17) {\rm ~MeV}.
\end{equation}
This is the most precise measurement of $f_{D^+}$~\cite{BESIII_f}.
The combined experimental error is 22\% while the LQCD error
reported at this conference is 10\% ~\cite{Wingate}. With the full
CLEO-c data sample a 2\% error for $f_{D^+}$ is expected. Similar
precision is expected for $f_{D_s^+}$ at $\sqrt{s}= 4140$~MeV.

\subsection{Measurement of the Charm Semileptonic Form Factors }

\begin{figure}[btp]
%\centerline{\epsfig{figure=kenu.eps,width=0.45\textwidth}}
\centerline{
\includegraphics[width=0.45\textwidth]{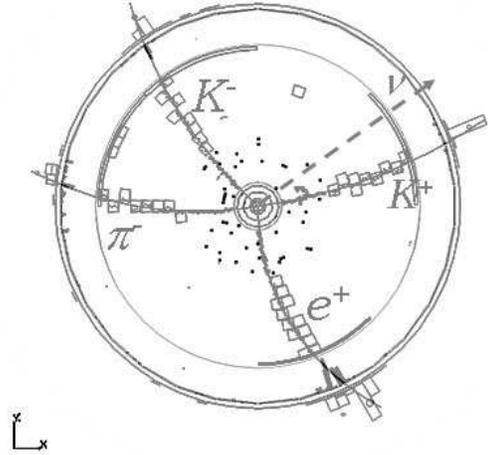}
} \vskip -1.0em \caption{ A CLEO-c event where $D^0 \rightarrow
K^- e^+ \nu_e, \bar{D^0}\rightarrow K^+ \pi^-$. }
\label{event_semilep} \vskip -1.5em
\end{figure}

\begin{figure*}[btp]
\centerline{
\epsfig{figure=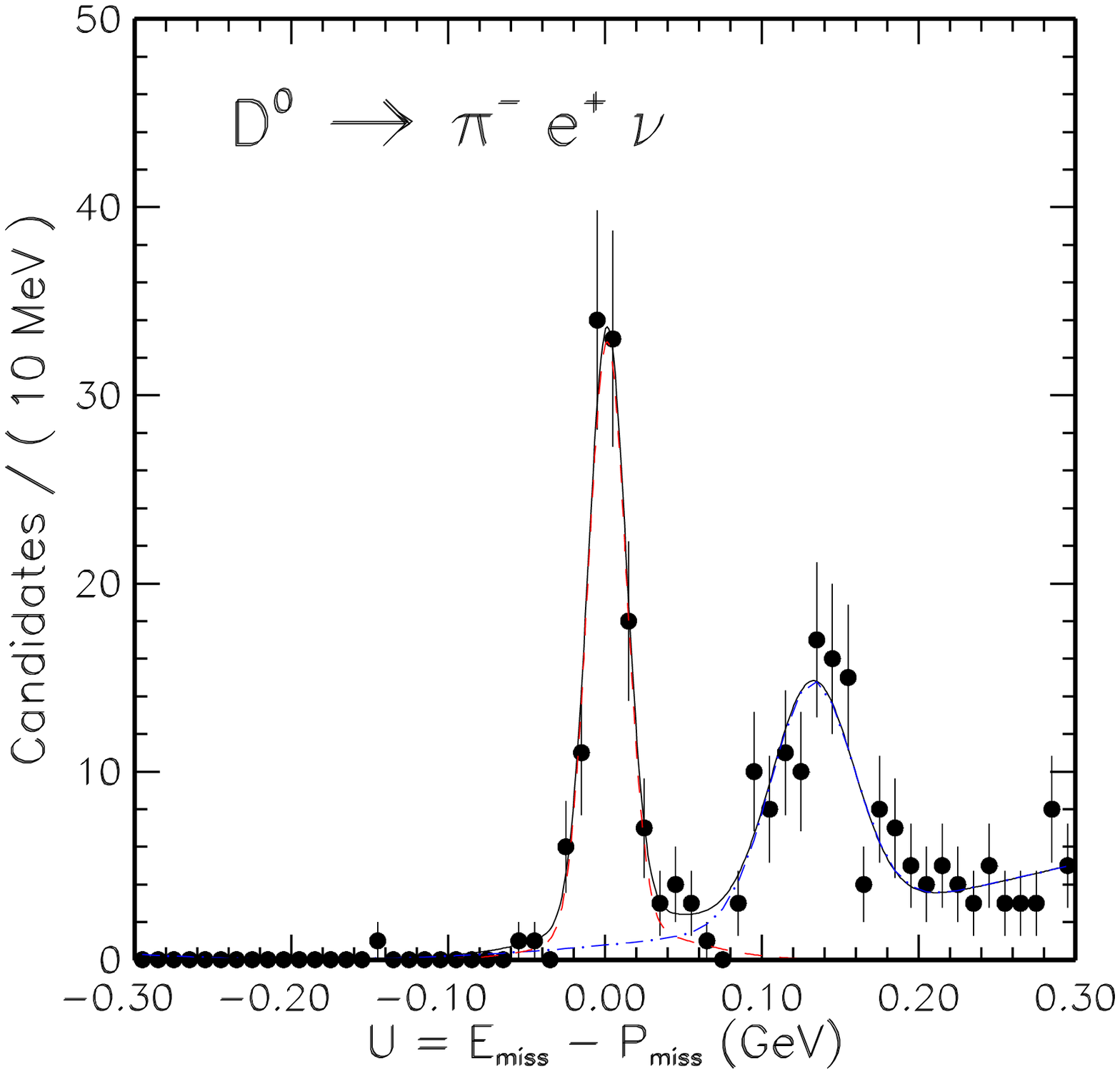,width=0.45\textwidth}
\hfill
\epsfig{figure=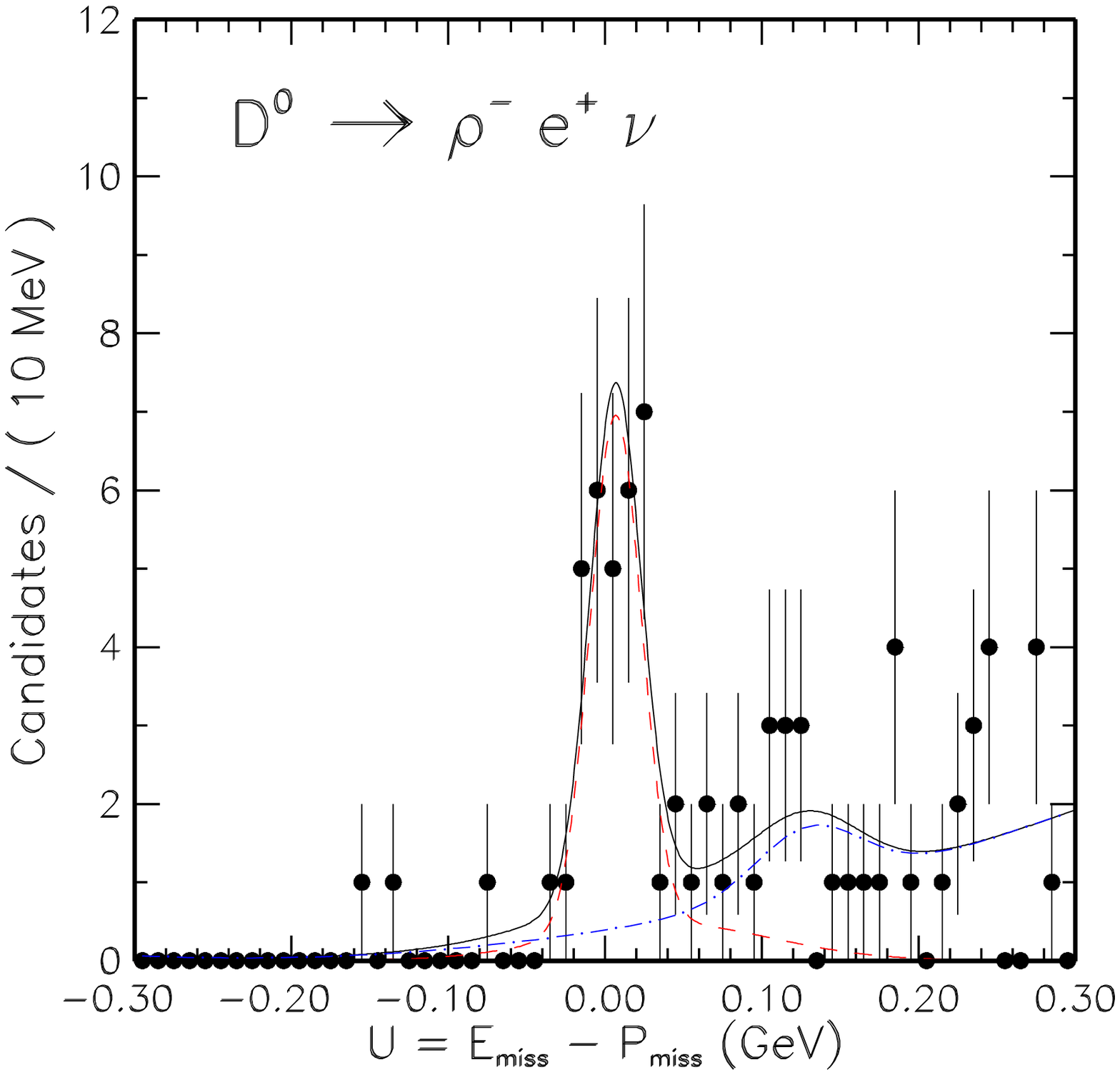,width=0.45\textwidth}
}
\vskip -1.0em
\caption{
The $U=E_{miss}-P_{miss}$ distribution in events with a
$\bar{D^0}$ tag, a positron, either
(a) a single charged track of the correct sign
or
(b) a $\rho^- \rightarrow \pi^+ \pi^0$, \emph{and}
and no additional (energetic) showers.
The peaks at zero and 0.13~GeV correspond to
(a) $D^0 \rightarrow \pi^- e^+ \nu_e$ and
$D^0 \rightarrow K^- e^+ \nu_e$
or
(b) $D^0 \rightarrow \rho^- e^+ \nu_e$ and
$ D^0 \rightarrow K^{*-} e^+ \nu_e$.
Preliminary. }
\label{fig:pi_rhoenu}
\end{figure*}

The measurement of semileptonic decays is also based on the use of
tagged events.
%where the cleanliness of the environment again
%provides nearly background-free signal samples.
A tagged event where the second $D$ decays semileptonically is
shown in Fig.~\ref{event_semilep}.
The analysis procedure, using
$D^0 \rightarrow \pi^- e^+ \nu_e$ as an example is as follows. A
positron and a hadronic track are identified recoiling against the
tag. The quantity $U = E_{miss}- P_{miss}$ is calculated, where
$E_{miss}$ and $P_{miss}$ are the missing energy and missing
momentum in the event. $U$ peaks at zero if only a neutrino is
missing.  The $U$ distribution in data is shown in
Fig.~\ref{fig:pi_rhoenu}(a) where a remarkably clean signal of about
100 events is observed for $D \rightarrow \pi e^+ \nu_e$.
The kinematic power of running at threshold also allows previously
unobserved modes such as $D^0 \rightarrow \rho^- e^+ \nu_e$ to be
easily identified see Fig.~\ref{fig:pi_rhoenu}(b).
CLEO-c results are given in Table~\ref{table:slbr}.
\begin{table}[tbp]
\caption{CLEO-c charm semileptonic branching ratios.
Further detail in Ref.~\cite{CLEO_semileptonic}.}
\label{table:slbr}
\begin{tabular}{@{}ll}
\hline
Mode                      & ${\cal B}$ (\%)  \\
\hline
$\dpie $                     &  $0.25 \pm 0.03 \pm 0.25 $ \\
$\dke $                    & $3.52 \pm 0.10 \pm 0.9$    \\
$\drhoe $                    &  $2.07 \pm 0.28 \pm 0.18$  \\
$ \dkste $                    & $0.19 \pm 0.04 \pm 0.02$   \\
 \hline
\end{tabular}
\vskip -1.5em
\end{table}
This modest data sample has
already produced the most precise determination of
${\cal B} (D^0 \rightarrow \pi^- e^+ \nu_e ) $.
With the full data set,
CLEO-c will make a significant improvement in the precision
with which each absolute charm semileptonic branching ratio is
known, see Table~\ref{table:slbr_proj}.
\begin{table}[tbp]
\caption{CLEO-c absolute semileptonic branching ratio projections.
(Some PDG2004 values are an average of $e$ and $\mu$.)
Further detail in Ref.~\cite{CLEO-cyellowbook}.}
 \label{table:slbr_proj}
\begin{tabular}{@{}lll}
\hline
Mode                      & \multicolumn{2} {c} { $ \delta {\cal B} / {\cal B} $ (\%) } \\
& PDG 2004 & CLEO-c  \\ \hline
$  \dke$ &5 &0.4 \\
$\dpie $  & 45  & 1.0   \\
$D^+ \rightarrow \pi^0 e^+ \nu_e  $                     &  48  &  2.0 \\
$D_s^+ \rightarrow \phi e^+ \nu_e $                    & 25   & 3.1   \\
 \hline
\end{tabular}
\vskip -1.5em
\end{table}

%Although the pilot run data sample is too small to determine the
%form factor in $D^0 \rightarrow \pi^- e^+ \nu_e$, it is useful to
%consider here how this will be done with a larger data sample.
%Energy-momentum conservation is used to determine $q^2$ in the
%laboratory frame
%\begin{eqnarray}
%E_W&=&E_{beam}-E_{had}, \non\\
%\vec{p}_W&=&-\vec{p}_{tag}-\vec{p}_{had},\non\\
%q^2&=&E_W^2-|\vec{p_W}|^2.
%\end{eqnarray}
%Here $E_W$ and $\vec p_W$ are the energy and momentum vector of
%the $\ell \nu$ system, equivalently the virtual $W$.
%$\vec{p}_{tag}$ is the momentum vector of the $D$ meson tag.
%$E_{had}$ and $\vec{p}_{had}$ are the energy and momentum vector
%of the hadronic system in the $D$ semileptonic decay.

The $q^2$ resolution is about 0.025 GeV$^{2}$, which is more than
a factor of 10 better than CLEO III which achieved a resolution of
0.4 GeV$^{2}$~\cite{Hsu}. This huge improvement is due to the
unique kinematics at the $\psi(3770)$ resonance, i.e. that the $D$
mesons are produced almost at rest. The combination of large
statistics, and excellent kinematics will enable the absolute
magnitudes and shapes of the form factors in every charm
semileptonic decay to be measured, in many cases to a precision of
a few per cent. This is a stringent test of LQCD.

By taking ratios of semileptonic and leptonic rates, CKM factors
can be eliminated. Two such ratios are ${\Gamma(D^+ \rightarrow
\pi^0 e^+ \nu_e)} / {\Gamma (D^+ \rightarrow \mu \nu_\mu)} $ and $
{\Gamma(D_s^+ \rightarrow (\eta {\rm ~or~} \phi ) e^+ \nu_e)} /
{\Gamma (D_s^+ \rightarrow \mu \nu_\mu)} $.
%\begin{eqnarray}
%{{\Gamma(D^+ \rightarrow \pi^0 e^+ \nu_e)} \over {\Gamma (D^+
%\rightarrow \mu \nu_\mu)} }, {\rm ~and~} {{\Gamma(D_s^+
%\rightarrow (\eta {\rm ~or~} \phi ) e^+ \nu_e)} \over {\Gamma
%(D_s^+ \rightarrow \mu \nu_\mu)} }
%\end{eqnarray}
These ratios depend purely on hadronic matrix elements and can be
determined to 4\% and so will test amplitudes at the 2\% level.
This is an exceptionally stringent test of LQCD.

If LQCD passes the experimental tests outlined above it will be
possible to use the LQCD calculation of the $B \rightarrow \pi$
form factor with confidence at the $B$ factories to extract a
precision $|\vub|$ from $B \rightarrow  \pi e^- \bar\nu_e$.  BaBar
and Belle will also be able to compare the LQCD prediction of the
shape of the $B \rightarrow \pi$ form factor to data as an
additional cross check.

Successfully passing the experimental tests will also allow CLEO-c
to use LQCD calculations of the charm semileptonic form factors to
directly measure $|\vcd|$ and $|\vcs|$, currently known to
7\% and 11\%~\cite{PDG2004}, with a greatly improved
precision of better than 2\% for each element. This in turn allows
new unitarity tests of the CKM matrix. For example, the second row
%of the CKM matrix can be tested at the 3\% level assuming theory
%can reach similar precision for semileptonic charm decays:
%of the CKM matrix can be tested at the 3\% level:
%\begin{equation}
%|\vcd|^2 + |\vcs|^2 + |\vcb|^2 = 1.
%\end{equation}
%The first column of the CKM matrix,
%\begin{equation}
%|\vud|^2 + |\vcd|^2 + |\vtd|^2 = 1,
%\end{equation}
%will be tested with similar precision to the first row (which is
%currently the most stringent test of CKM unitarity). Finally, the
%ratio of the long sides of the $uc$ unitarity triangle will be
%tested to 1.3\%.
of the CKM matrix can be tested at the 3\% level;
the first column of the CKM matrix
will be tested with similar precision to the first row (which is
currently the most stringent test of CKM unitarity); finally, the
ratio of the long sides of the $uc$ unitarity triangle will be
tested to 1.3\%.

Table~\ref{table:combine} provides a summary of projections for
the precision with which  the CKM matrix elements will be
determined if LQCD passes the CLEO-c tests in the $D$ system.  In
the tabulation the current precision of the CKM matrix elements is
obtained by considering methods applicable to LQCD, for example
the determination of $|\vcb|$ and $|\vub|$ from inclusive decays
and OPE is not included. The projections are made assuming $B$
factory data samples of 500~fb$^{-1}$ and improvement in the
direct measurement of $|\vtb|$ expected from the Tevatron
experiments~\cite{Swain}.
%(Figure~\ref{CKM}(b) was made
%assuming the precision given in Table~\ref{table:combine}.)

\begin{table}[tbp]
\caption{LQCD impact (in per cent) on the precision of CKM matrix elements.
Further detail in Ref.~\cite{CLEO-cyellowbook}.}
 \label{table:combine}
\centering
\begin{tabular}{lrrrrrr}
\hline
% ~~ & $|\vcd|$ & $|\vcs|$ & $|\vcb|$ & $|\vub|$ & $|\vtd|$ & $|\vts|$ \\
%2004  & 7\% & 11\% & 4\% & 15\% & 36\% & 39\%   \\
%LQCD & 1.7\% & 1.6\% & 3\% & 5\% & 5\% & 5\% \\
  ~~ & $\vcd$ & $\vcs$ & $\vcb$ & $\vub$ & $\vtd$ & $\vts$ \\
\hline
2004  & 7 & 11 & 4 & 15 & 36 & 39   \\
LQCD & 1.7 & 1.6 & 3 & 5 & 5 & 5 \\
 \hline
\end{tabular}
\vskip -1.5em
\end{table}

\subsection{Probing QCD with Heavy Quarkonia}

Here the twin goals are to verify the theoretical tools for
strongly coupled field theories and quantify the accuracy  for
application to flavor physics.  As the same actions are used in
both onia and $B/D$ calculations, onia provide an independent
calibration of $c$ and $b$ quark actions used in $B/D$ physics.
Heavy quarkonia is the richest calibration/testing ground for
lattice techniques.

In the $\psi$ and $\Upsilon$ systems there are more than thirty
gold plated (few \%) lattice calculations now possible.
Measurements of masses and spin fine structure for $S$, $P$, and $D$
states reveal the magnitude of relativistic corrections and the
nature of confinement. The measurement of leptonic widths for $S$
states test wave function techniques that are important for
calculating decay constants, while electromagnetic transitions for
$P \rightarrow S$ and $S \rightarrow P$ matrix elements are
related to calculations of semileptonic form factors.

Recently, there has been an order of magnitude increase in the
data available to test predictions; Upsilonia at CLEO III and
charmonium at BES II and CLEO-c. One noteworthy discovery has been
the observation of the $1^3D_J$ states. The $b\bar{b}$ system is
unique as it has states with $L = 2$ that lie below the open-flavor
threshold.
These states are of considerable theoretical
interest~\cite{bib:EFI01-14}.
The mass of the $\Upsilon(1^3D_2)$
tests the lattice at large~$L$.  CLEO has observed the
$\Upsilon(1^3D_2)$ state in the four-photon cascade $\Upsilon(3S)
\rightarrow \gamma_1\chi'_b \rightarrow \gamma_1\gamma_2
\Upsilon(^3D_J) \rightarrow \gamma_1\gamma_2\gamma_3 \chi_b
\rightarrow \gamma_1\gamma_2\gamma_3\gamma_4\ell^+\ell^-$,
finding~\cite{Upsilon1D}
$M(\Upsilon(1^3D_2)) = (10161.1 \pm 0.6 \pm 1.6)$~MeV/$c^2$,
in good agreement
with~\cite{Davies}. Some other important goals are the observation
of the $\eta_b$ and $h_b$ in the $\Upsilon$ system and the $h_c$
in the charmonium system.

\subsection {Glueballs and hybrid states}

QCD is the only known theory in nature where gauge particles can
also be constitutents. Glueballs and hybrids are fundamental
states of the theory and the current lack of strong, unambiguous
evidence for their existence is a challenge to QCD. If glueballs
are observed this will be a major discovery in particle physics
and a highly nontrivial test of lattice QCD~\cite{Morningstar}.
The approximately one billion $J/\psi$ produced at CLEO-c will be
a glue factory to search for glueballs and other glue-rich states
via $J/\psi \rightarrow \gamma gg \rightarrow \gamma X$ decays.
The region  $1 < M_X < 3$~GeV/$c^2$ will be explored with partial
wave analyses for evidence of scalar or tensor glueballs,
glueball-$q\bar{q}$ mixtures, quark-glue hybrids and other new
forms of matter. The goals include the establishment of masses,
widths, spin-parity quantum numbers, decay modes and production
mechanisms for any identified states, a detailed exploration of
reported glueball candidates such as
%such as the tensor candidate $f_J(2220)$
the scalar states $f_0(1370)$, $f_0(1500)$ and $f_0(1710)$, and
the examination of the inclusive photon spectrum $J/\psi
\rightarrow \gamma$X with $<$ 20 MeV photon resolution and
identification of states with up to 100 MeV width and inclusive
branching ratios above $1 \times 10^{-4}$.

\section{ THE EXPERIMENTER'S VIEW}

\subsection{The bottom line}

How can we be sure that if LQCD works for $D$ mesons it will work
for $B$ mesons? Or, equivalently, are CLEO-c data sufficient to
demonstrate that lattice systematic errors are under control?
There are a number of reasons to answer this question in the
affirmative. (1) There are two independent effective field
theories: NRQCD and the Fermilab method. (2) The CLEO-c data
provide many independent tests in the $D$ system; leptonic decay
rates, and semileptonic modes with rate and shape information. (3)
The $B$ factory data provide additional independent cross checks
such as $ { d \Gamma(B \rightarrow \pi \ell \nu ) / d p_\pi}$. (4)
Unlike models, methods used for the $D/B$ system can be tested in
heavy onia with measurements of masses, and mass splittings,
$\Gamma_{ee}$ and electromagnetic transitions. (5) The main
systematic errors limiting accuracy in the $D/B$ systems are:
chiral extrapolations in $m {\rm light}$, perturbation
theory, and finite lattice spacing. These are similar for
charm and beauty quarks. In my opinion a combination of CLEO-c
data in the $D$ systems and onia, plus information on the light
quark hadron spectrum, can clearly establish whether or not
lattice systematic errors are under control.

While this picture is encouraging, experimentalists also have
concerns. The lattice technique is all encompassing but LQCD
practitioners are very conservative about what can be calculated.
Much of the excitement this summer in flavor physics revolves
around whether $\sin 2\beta (\psi K_S^0) = \sin 2 \beta (\phi
K_S^0 )$, and also the observation of $A_{CP}$ in $B \rightarrow K
\pi$~\cite{ACP}. The lattice is not yet able to contribute in
these areas. There is a need to move beyond gold-plated quantities
in the next few years: for example resonances such as $\rho, ~\phi
{\rm ~and~} K^*$ may be difficult to treat on the lattice, but they
feature in many important $D$ semileptonic decays which will be
well measured by CLEO-c. There is also a pressing need to be able
to calculate for states near threshold such as $\psi(2S)$ and
$D_s(0)^+$, and hadronic weak decays as well.

\subsection {Systematic Errors}

It will take accurate and precise experimental measurements
combined with accurate and precise theoretical calculations to
search for new physics in the CKM matrix. Therefore, it is
essential to chase down each and every source of systematic error
in lattice calculations.

Usually, by far the most demanding part of a precision
experimental measurement is the careful evaluation of the
systematic errors. Therefore one way an experimenter evaluates the
quality of a measurement is by the completeness of the systematic
error analysis.  As lattice results increase in precision,
experimentalists will expect to see full error reporting and
discussion of errors with every lattice calculation. So, with
experimentalists, phenomenologists, and lattice colleagues in mind
lattice results should:
\begin{enumerate}
\item Include a comprehensive table of systematic errors with {\em
every} calculation. Many calculations already have this.  An error
budget makes it more straightforward to compare results from
different groups. It is understood that different methods will
have somewhat different lists.
\item Include a statement of whether an error is Gaussian or
non-Gaussian. Errors are often estimates of higher order terms in
a truncated expansion, so the quoted error bar is non-Gaussian.
For the statistical error a distribution could be provided.
\item Report the correlation between individual sources of
systematic error (if such correlation exists).
\item Provide a total systematic error by suitably combining
individual errors. This is redundant and should not replace the
individual error breakdown, but certainly convenient.
\end{enumerate}

\subsection{Outlook}

I will begin this section with a few quotes that summarize the
outlook over the next few years.

``Expect to see a growing number of lattice results for gold
plated quantities within the next few years with an ultimate goal
of a few \% errors within five years."  A prominent lattice
theorist (2003).

``Prediction is better than postdiction." Every experimentalist
(every time).

``We need high precision experimental results in order to test
lattice QCD, we need CLEO-c for $D$ decays." A prominent lattice
theorist (2003).

``CLEO-c may have a few \% preliminary determination of $f_{D^+}$
as early as the summer conferences in 2005." A CLEO-c
collaboration member (summer, 2004).

A more precise unquenched lattice calculation of $f_{D^+}$ with
complete error report {\em before} the CLEO-c result from the
first full run is announced next summer, will clearly demonstrate
the {\em current} precision of the lattice approach to the
community and give credibility to the goal of a few \% errors.

A similar argument applies to the calculation of the form factors
in $D \rightarrow K/ \pi e \nu_e$ by summer 2005 and $f_{D_s^+}$
and form factors in $D_s^+$ semileptonic decays by summer 2006. It
must be noted, however, that the precision of the CLEO-c results,
and when that precision is reached, depend crucially on the
luminosity performance of CESR-c.

\section{SUMMARY}

This is a special time in flavor physics. The lattice goal is to
calculate to a few percent precision in the $D, B, \Upsilon, {\rm
~and~} \psi$ systems. CLEO-c, and later BES III, is about to
provide few per cent precision tests of lattice calculations in
the $D$ system and in heavy onia, which will quantify the accuracy
for the application of LQCD to the B system.  Then BaBar, Belle,
CDF, D0, BTeV, CMS, ATLAS, and LHC-b data, in combination with
LQCD will lead to a few per cent determinations of $|\vub|,
|\vcb|, |\vtd|,$ and $|\vts|$.

To borrow from the title of Ref.~\cite{Davies}: precision LQCD
confronts experiment, but equally, precision experiment confronts
LQCD. The combination of LQCD and CLEO-c have the potential to
maximize the sensitivity of the flavor physics program to new
physics this decade and pave the way for understanding beyond the
SM physics at the LHC in the coming decade.

\section{ACKNOWLEDGEMENTS}

I thank my colleagues on the CLEO Collaboration for allowing me to
include many CLEO results in this talk. I thank Elisabetta
Barberio, Tom Browder, Toru Ijima and Bruce Yabsley (Belle), David
Hitlin, Jeff Richman, Brian Meadows, and David Williams (BaBar),
Daniela Bortoletto, Art Garfinkel and Stefano Giagu (CDF), Vivek
Jain, John Womersley, and Daria Zieminska (D0) and Ed Blucher
(KTeV) for making results and projections from their respective
experiments available. I am grateful to the following % colleagues
for valuable discussions on the relationship between the lattice
and CLEO-c: Nora Brambilla, Aida El-Khadra, Andreas Kronfeld,
Peter Lepage, Vittorio Lubicz, Paul Mackenzie, Roberto Mussa, Jim
Napolitano, Matthias Neubert, Jon Rosner, Anders Ryd, Kam Seth,
Jim Simone, Sheldon Stone, Matt Wingate, and Jim Wiss. Finally, thanks to the
organizers of Lattice 2004 for a stimulating and superbly run
conference.

\end{document}